# Gamma and Decay Energy Spectroscopy Measurements of Trinitite

David J. Mercer,[a] Katrina E. Koehler,[a] Mark P. Croce,[a] Andrew S. Hoover,[a] Philip A. Hypes,[a] Stosh A. Kozimor,[a] Veronika Mocko,[a] Paul R. J. Saey,[b] Daniel R. Schmidt,[c] and Joel N. Ullom[c]

[a]*Los Alamos National Laboratory, Los Alamos, New Mexico*
[b]*International Atomic Energy Agency, Division of Safeguards, Vienna, Austria*
[c]*National Institute of Standards and Technology, Boulder, Colorado*

**Abstract**: We report gamma ray spectroscopy measurements of trinitite samples and analogous samples obtained from detonation sites in Nevada and Semipalatinsk, as well as *in situ* measurements of topsoil at the Trinity site. We also report the first isotopic composition measurements of trinitite using the novel forensics technique of decay energy spectroscopy (DES) as a complement to traditional forensics techniques. Our measurements are compared to other published results.

## I. Introduction

The world's first nuclear device was detonated on 16 July 1945 at the United States Army Air Forces (USAAF) Alamogordo Bombing and Gunnery Range southeast of Socorro, New Mexico. "The Gadget" was placed on a 100-foot steel tower and detonated with a yield of about 21 kilotons; a new analysis by Selby et al.[1] appearing in this issue gives 24.8±2.0 kt. The blast produced a circular green glassy layer of material on the ground about 370 m in diameter and centered on the tower. This is the anthropogenic mineral "trinitite." Early photos show a surface that is nearly completely covered with this material. The surface was bulldozed in 1952, partially in response to scavenging by collectors. Most of what remains near ground zero today appears to be neutron-irradiated soil and only trace amounts of glassy trinitite; however, some areas rarely accessible to the public remain a rich trove (see Figure 1). Some trinitite remains in the subsoil, and small beads may be brought to the surface by ants.[2]

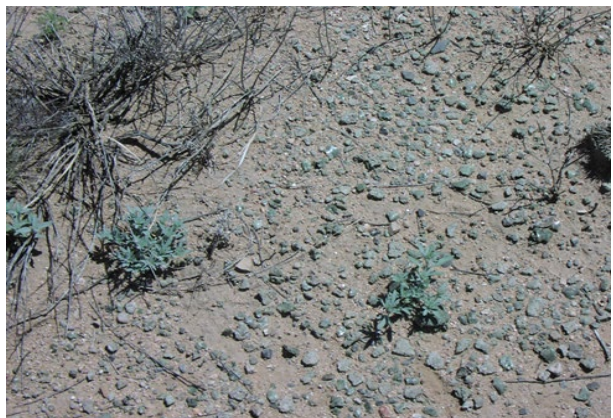

Figure 1. A relatively undisturbed patch of ground showing Trinitite fragments, courtesy of Robert Hermes.

Trinitite is known to contain fission products, neutron activation products, and residual nuclear fuel, which are detectable even today, as reported in prior studies.[3–7] Some studies have explored the relationships between radionuclide activities in samples and their collection points,[6,8] although we do not attempt this level of detail. Microscopic mineralogical studies indicate that the most radioactive components of trinitite consist of soil that was drawn into the fireball, vaporized and mixed with explosion debris, then condensed and rained down rather than forming directly on the ground.[9–12] The soil itself was also exposed to a neutron flux resulting in activation products. We performed gamma-ray measurements of trinitite, related mineral samples, and Trinity site soil; preliminary results were published in Mercer et al.[13] Decay Energy Spectroscopy (DES) measurements require a very small amount of material (~1 Bq), so a small aerodynamic bead of trinitite was used for DES.

## II. In situ soil measurements

A visit to the site in August 2010 presented an opportunity to collect high-resolution gamma ray spectra of Trinity site soil to quantify the presence of residual radioactive isotopes. The visit was out of scientific curiosity and not an official function of the IAEA or LANL. The instruments chosen were typical for IAEA inspectors. We performed six measurements of the soil surface in various locations using an HPGe-based Ametek® Ortec® Micro-Detective. Non-random locations were selected where strong activity was identified in a cursory survey using a NaI-based Thermo identi-FINDER™ (HM5). Most locations were northeast of ground zero, corresponding to the downwind direction at the time of the original detonation. Figure 2 shows a typical measurement in progress; note that the surface here lacks any visible trinitite fragments.

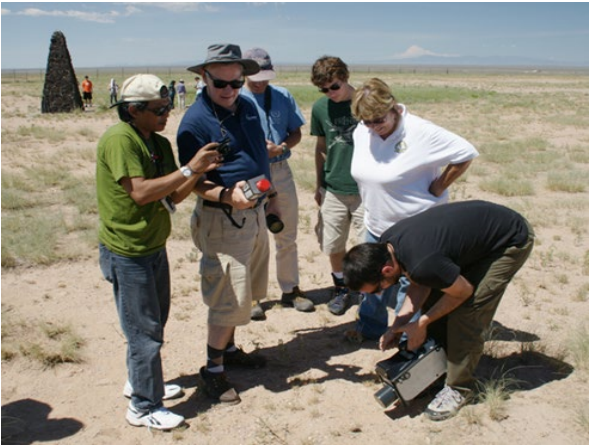

Figure 2. An *in situ* measurement; the obelisk marks ground zero.

The soil spectra are dominated by $^{152}$Eu, as shown in Figure 3. $^{137}$Cs and $^{154}$Eu are also visible in all soil spectra, and $^{60}$Co is visible in some of them. The radioisotopes $^{152}$Eu ($t_{1/2}$ = 13.5 years) and $^{154}$Eu ($t_{1/2}$ = 8.5 years) are from neutron irradiation of natural europium in the soil, $^{137}$Cs ($t_{1/2}$ = 30.1 years) is a fission product from the detonation, and $^{60}$Co ($t_{1/2}$=5.3 years) was produced mainly by activation of steel from the tower. Many shorter-lived fission and activation radionuclides were likely produced but are no longer visible. We observed 0.0 – 0.4 Bq/cm² of $^{60}$Co, 1.0 – 7.4 Bq/cm² of $^{137}$Cs, 7.8 – 37.6 Bq/cm² of $^{152}$Eu, and 0.1 – 3.4 Bq/cm² of $^{154}$Eu at the various locations.

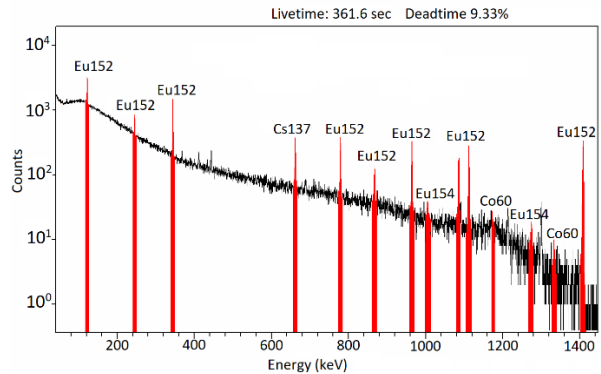

Figure 3. Typical gamma ray spectrum of Trinity Site soil showing signature photopeaks from $^{60}$Co, $^{137}$Cs, $^{152}$Eu, and $^{154}$Eu.

Surprisingly, no signature of $^{241}$Am ($t_{1/2}$=433 years) was visible in any of our *in situ* soil spectra. Most published measurements of mineral trinitite samples, including our own, report strong visibility of $^{241}$Am, so we had fully expected to see a residue in the soil. The $^{241}$Am in trinitite samples is due mostly to decay of $^{241}$Pu ($t_{1/2}$=14.35 years) that was part of the original fuel and produced by neutron capture during the detonation. It is likely that the soil contains very little actual trinitite and consists mostly of neutron-irradiated subsoil. Most americium apparently was removed removed along with the top layer of soil during surface remediation or leached deeper into the soil by rain; what remains near the surface is below our limit of detection. The $^{137}$Cs / $^{152}$Eu activity ratio measured in the soil is a factor of 20 smaller than typically observed in samples of trinitite, suggesting that most of the $^{137}$Cs has been removed or leached, and $^{60}$Co is also severely reduced. Other publications also report soil measurements, but with somewhat higher relative $^{60}$Co levels.[6,14]

## III. Laboratory measurements of Trinitite

Gamma spectroscopy measurements were made in a laboratory environment of trinitite and analogous samples. This allowed us to quantify the radioisotope activity and compare to previously published results and to ratios from our *in situ* soil measurements. We did not collect any soil samples for laboratory gamma spectroscopy measurements.

We measured two samples of green trinitite. Sample "A" was collected from the Trinity site by a museum curator and consists of 125.6 grams of typical green glass pieces. Sample "B" is a 248.6-gram portion of a bulk quantity borrowed from Los Alamos National Laboratory. It is similar in appearance to Sample A. We measured 21.2 grams of rare red trinitite provided by a Los Alamos collector. We also measured a sample from a low-yield Nevada Test Site detonation "Johnnie Boy" (11 July 1962). This is not precisely "trinitite" so the generic mineral term "atomsite" is more appropriate. It is a 4.4-gram piece, shiny black and porous on one side, dull grey on the other, about 2 cm in diameter. Kharitonchik is the Soviet analog to trinitite, found at the Semipalatinsk Test Site (STS) in Kazakhstan. We measured a 42.3-gram sample consisting of a single piece about 5 cm in diameter that is reportedly from the first Soviet test (29 August 1949); this was collected by Byron Ristvet with the permission of the Kazakhstan Nation Nuclear Center and the Russian Federation Atomic Energy Agency, October 2006. Figure 4 shows some of the samples. All of our laboratory gamma spectroscopy measurements were completed in June 2011.

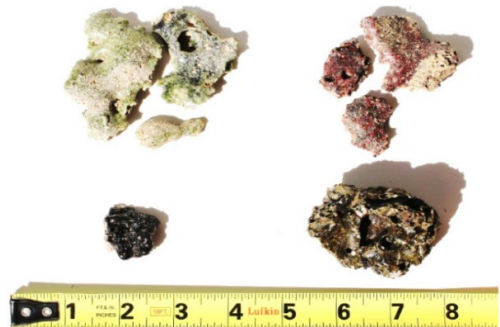

Figure 4. Top: example green and red Trinitite measured in this project; bottom left: Nevada atomsite; bottom right: Kharitonchik.

A long-count spectrum from Sample "B" appears in Figures 5 and 6. In contrast to the soil spectra, radionuclides $^{241}$Am, $^{133}$Ba ($t_{1/2}$=10.5 years), and $^{239}$Pu ($t_{1/2}$=24,100 years) are evident. $^{239}$Pu was the primary nuclear fuel, and $^{133}$Ba is mostly from $^{132}$Ba($n,\gamma$) activation in baratol, a type of explosive used in the device that is approximately 70% barium nitrate. Additional radionuclides visible in our Kharitonchik sample include $^{155}$Eu ($t_{1/2}$ = 5.0 years) which is a fission product, and $^{237}$Np ($t_{1/2}$ = 2,140,000 years), which is evident from its $^{233}$Pa daughter.

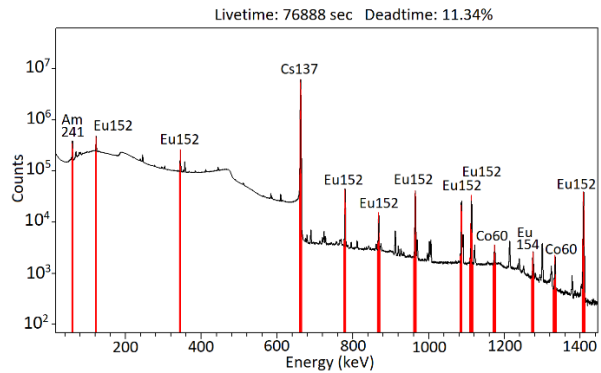

Figure 5. $^{60}$Co, $^{137}$Cs, $^{152}$Eu, $^{154}$Eu, and $^{241}$Am seen in Sample "B".

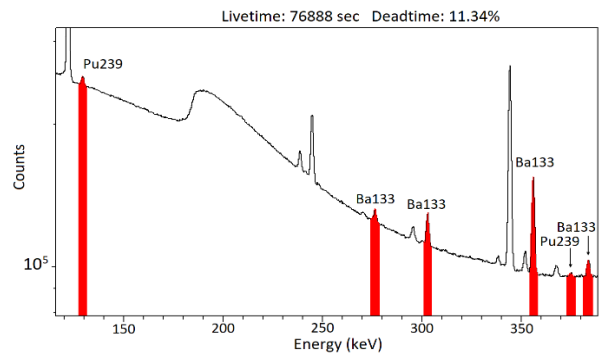

Figure 6. Detail from Sample "B" showing $^{133}$Ba and $^{239}$Pu.

Gamma quantification measurements used an Ortec Detective-EX-100. Samples were counted in a close configuration in a low-background laboratory without additional shielding or collimation. Absolute calibration is based on a secondary measurement of each sample at 25 cm (to reduce geometric calibration errors). Our results are shown in Figure 7, where activities have been decay-corrected to reflect activities at the time of detonation. For $^{241}$Pu the calculated activity is based on the assumption that it is the sole source of the $^{241}$Am observed today, consistent with Pittauerová et al.;[6] no $^{241}$Pu is directly observed in our gamma measurements. DES measurements are sensitive to beta-emitters such as $^{241}$Pu; however, discrimination and quantification are challenging, and we did not attempt to do so.

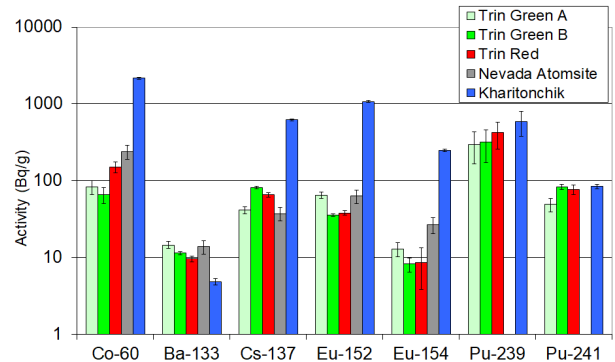

Figure 7. Comparison between results from our five samples (logarithmic scale). No $^{239}$Pu or $^{241}$Am are observed in atomsite.

Our three trinitite samples show some variation in activities, which may be due to location, uneven mixing, and variations of neutron fluence. A notable difference among the samples is the two-times higher activity of $^{60}$Co visible in the red trinitite. $^{60}$Co is a product of the $^{59}$Co($n,\gamma$) reaction and cobalt is a component of steel. Presumably, our red sample contained more steel residue than the green. Others have observed that red trinitite contains more Co, Ni, Cu, and Pb than green samples.[15] Our green Co, Ba, Cs, and Eu results are consistent with Refs. 3–6 except that we are not able to observe or quantify $^{155}$Eu in contrast to Pittauerová et al..[6] We do not know of other published measurements of red trinitite activities.

$^{239}$Pu was observed in our three trinitite samples but is difficult to quantify due to the weak signature and interference from other gamma rays. Our reported activities are roughly four times higher than in Parekh et al.[5] or Pittauerová et al..[6] Radiochemical techniques (see, for example, Refs. 16−19) or DES can be far more sensitive for measuring Pu in trinitite

The Nevada atomsite sample showed radionuclide activities generally similar to green trinitite, except for 2× higher activity of $^{60}$Co and absence of detectable $^{239}$Pu or $^{241}$Am (which was plainly visible in all of our other samples). From this absence we infer that the fuel for this detonation did not contain plutonium. This conclusion is supported by a US Government corrective action investigation plan, which states that "plutonium and americium are not expected as a result of the test at Johnnie Boy but could be present at low levels as the result of tests nearby."[20]

$^{233}$Pa from $^{237}$Np decay is tentatively observed in our Kharitonchik sample and is also reported in published STS soil measurements.[17,18] The $^{237}$Np is generated by $^{238}$U($n,2n$) interactions or multiple neutron capture on $^{235}$U, producing $^{237}$U ($t_{1/2}$ = 6.75 days) which decays to $^{237}$Np. The $^{233}$Pa daughter has a short half-live (27.4 days) but is continuously replenished by $^{237}$Np decay.

The Kharitonchik sample displays order-of-magnitude higher activities of $^{60}$Co, $^{137}$Cs, $^{152}$Eu, and $^{154}$Eu than seen in trinitite, in agreement with qualitative observations of Pittauerová et al..[6] We are not aware of any other published quantitative activities, but we can compare to the soil samples of Yamamoto et al.[17] collected at STS near ground zero of the first Soviet test. There is no record of surface remediation, so the STS soil might contain Kharitonchik fragments. Our $^{60}$Co and $^{152}$Eu results are similar to Yamamoto et al.,[17] our $^{137}$Cs and $^{154}$Eu results are 2.5× higher, $^{241}$Pu (from $^{241}$Am) is 4.5× higher, our $^{239}$Pu and $^{237}$Np results are an order of magnitude higher, and the reference does not report $^{133}$Ba or $^{155}$Eu. We are not aware of any other published detection of $^{133}$Ba in Kharitonchik or STS soil, although its absence is noted and discussed with some confusion in Schlauf et al.[4] and Pittauerová et al..[6] At the level observed we cannot conclude whether the $^{133}$Ba is from irradiation of device components or of natural barium in the soil. See Table I for a comparison between our gamma measurement results and those from other publications.

While gamma spectroscopy is able to detect $^{239}$Pu in some of the mineral samples, the relevant gamma rays have intensities of only $10^{-4} - 10^{-5}$ per decay, so sensitivity is poor. Detection of $^{240}$Pu, with its even weaker gamma signature confounded by overlapping peaks, is not possible, so the plutonium isotopic ratio is elusive. We turn to DES with its higher sensitivity.

### IV. Plutonium Isotopic Composition

Determining the plutonium isotopic composition of nuclear detonation debris is an important forensics goal. Traditional forensics methodologies include alpha, gamma, and mass spectrometry, often with a step of chemical purification or laser ablation. Each of these approaches has advantages and disadvantages, leading to a combination of approaches for a thorough forensics analysis.

Table I: Concentrations of Radionuclides Observed in Trinitite[a]

| Radio-nuclide | Trinitite Green A (This Work) | Trinitite Green B (This Work) | Trinitite Red (This Work) | Trinitite Atkatz 1995[3] | Trinitite Schlauf 1997[4] | Trinitite Parekh 2006[5,c] | Trinitite Pittauerová 2010[6,d] | Trinitite Bellucci 2013[8,e] |
|---|---|---|---|---|---|---|---|---|
| $^{60}$Co | 82.9 ± 17.3 | 65.7 ± 15.1 | 150.8 ± 24.7 | – | 44 ± 4 | 44.4 – 62.0 | 47.7 ± 4.6 | – |
| $^{133}$Ba | 14.5 ± 1.5 | 11.3 ± 0.6 | 9.6 ± 0.8 | – | 9.9 ± 0.6 | 7.6 – 9.8 | 9.10 ± 0.77 | 27 – 110 |
| $^{137}$Cs | 41.4 ± 4.2 | 81.1 ± 3.1 | 65.6 ± 4.4 | 83.2 | 90 ± 9 | 27.33 – 121.8 | 48.3 ± 1.4 | 10 – 371 |
| $^{152}$Eu | 64.6 ± 6.5 | 35.7 ± 1.4 | 37.9 ± 2.6 | – | 27 ± 1 | 22.6 – 78.9 | 26.0 ± 1.1 | 8 – 113 |
| $^{154}$Eu | 12.9 ± 2.6 | 8.2 ± 1.7 | 8.6 ± 4.8 | – | 4.8 ± 0.6 | 2.5 – 16.1 | 7.08 ± 0.24 | 29 – 507 |
| $^{155}$Eu | – | – | – | – | – | – | 274 ± 25 | 0.2 – 1.5[f] |
| $^{237}$Np | – | – | – | – | – | – | – | – |
| $^{239}$Pu | 296 ± 132 | 317 ± 144 | 420 ± 161 | – | – | 86.3 ± 2.7 | 73.6 ± 3.5 | 6,710-13,753[f] |
| $^{241}$Am[b] | 1.5 ± 0.3 | 2.5 ± 0.2 | 2.3 ± 0.3 | – | 2.9 ± 0.5 | 1.84 – 4.14 | 1.87 ± 0.08 | 1 – 37 |
| $^{241}$Pu[b] | 49.1 ± 10.0 | 82.9 ± 6.6 | 76.9 ± 10.6 | – | 100 ± 17 | 63 – 142 | 63.6 ± 2.7 | 34 – 1250 |

| Radio-nuclide | Trintity Site Soil (This Work)[g] | Trintity Site Soil Hansen 1985[14] | Trintity Site Soil Pittauerová[6] | Nevada Atomsite (This Work) | STS Kharitonchik (This Work) | STS Soil Yamamoto 1996[17,h] | STS Soil Beasley 1998[18,h,i] |
|---|---|---|---|---|---|---|---|
| $^{60}$Co | 0.0 – 0.4 | 64 – 320 | 2.9 – 20.8 | 239.2 ± 48.4 | 2162.4 ± 70.1 | 2033.4 ± 45.1 | – |
| $^{133}$Ba | – | – | 0.73 – 1.73 | 0.10 – 0.36 | 13.8 ± 2.8 | 4.8 ± 0.5 | – |
| $^{137}$Cs | 1.0 – 7.4 | 0.38 – 1.87 | 0.14 – 0.88 | 37.0 ± 7.4 | 621.1 ± 19.7 | 235.5 ± 2.0 | – |
| $^{152}$Eu | 7.8 – 37.6 | 3.2 – 347 | 7.9 – 33.4 | 63.0 ± 12.6 | 1071.2 ± 34.0 | 970.3 ± 10.1 | – |
| $^{154}$Eu | 0.1 – 3.4 | – | 1.1 – 4.8 | 26.5 ± 6.3 | 247.5 ± 8.3 | 110.2 ± 4.9 | – |
| $^{155}$Eu | – | – | 0 – 10.8 | – | 3347 ± 1677 | – | – |
| $^{237}$Np | – | – | – | – | 0.72 ± 0.03 | 0.076 ± 0.003 | 2.7×10$^{-5}$ ± 2×10$^{-6}$ |
| $^{239}$Pu | – | – | – | – | 586 ± 211 | 27.9 ± 0.4 | 0.134 ± 0.004 |
| $^{241}$Am[b] | – | – | 0.008 – 0.037 | – | 2.4 ± 0.2 | 0.52 ± 0.01 | 0.0770 ± 0.0001 |
| $^{241}$Pu[b] | – | – | 0.28 – 1.25 | – | 83.6 ± 6.1 | 18.5 ± 0.4 | 2.737 ± 0.003 |

[a] Except where noted, concentrations are in Bq of radionuclide per gram of sample, decay-corrected to the date of detonation. Uncertainties are 1σ.
[b] Activity for $^{241}$Am is reported per the measurement date with no decay corrections. Activity for $^{241}$Pu is computed from these $^{241}$Am observations and is decay-corrected to the date of detonation.
[c] Maximum and minimum values are reported for three samples.
[d] Median value is reported for 11 samples.
[e] Maximum and minimum nonzero values are reported for 49 samples. Zero activity was reported for at least one sample for each radionuclide.
[f] Reference reports observation of $^{155}$Eu in four samples and $^{239}$Pu in three samples but we find the quantitative results hard to understand.
[g] Soil measurements from the present work are reported in reported in Bq/cm$^2$ rather than Bq/g so the values are not directly comparable to other works.
[h] Radiochemical techniques were applied to extract actinide activities. Combined activity of $^{239}$Pu+$^{240}$Pu is reported in these references with no decay correction.
[i] STS soil sample was collected 1.1 kilometers from ground zero.

In standard alpha spectroscopy with energy resolution of 15–20 keV, the 5.486 MeV alpha from [241]Am cannot be resolved from the 5.499 MeV alpha from [238]Pu. Because of interferences like this, chemical separation is done before source preparation in order to resolve the two radioisotopes. To avoid energy loss within the sample, alpha sources must be thin and uniform, which often means electrodeposition if the sample size is small.[21]

Mass spectrometry yields high-precision and high-accuracy isotopic ratios. Mass spectrometry works by ionizing a small sample and then using mass-to-charge ratio to separate the ions. This method typically requires chemical separations to avoid the isobaric interferences, for example, between [241]Am and [241]Pu.[21]

The radiochemical preparation for both alpha spectroscopy and mass spectrometry can be time consuming, delaying results in a timely forensics investigation. Further, the chemical process serves to homogenize the sample, averaging out any spatial information within the sample.[6,7,22] As a result, complementary forensics measurements that do not require chemical separation and utilize smaller sample sizes could be valuable.

The first measurement of the plutonium isotopic ratio in trinitite was made by Anderson and Sugarman in 1945, described in Ref. 23 and summarized in Ref. 19. Using a combination of alpha and fission counting on a chemically purified sample, they deduced a [240]Pu/[239]Pu mass ratio of 0.018.

As part of a 1978 environmental study, Douglas combined alpha and mass spectrometry measurements of three purified samples to measure plutonium isotopics.[24] His results span a wide range, and our average of his 22 measurements is 0.027±0.007 with uncertainty calculated by the standard deviation of his results. The ratios that we report here and hereafter may be assumed to be [240]Pu/[239]Pu mass ratios observed in trinitite, decay-correction to the date of the detonation, with 1$\sigma$ uncertainties. In most cases the uncertainty reported includes only the statistical, not systematic, components.

A frequently cited isotopic measurement by Parekh et al.[5] was done in 2006 using a combination of alpha and gamma spectroscopy. In the radiochemical preparation, the aggregate 5.4 g of trinitite was milled and then a 0.5 g aliquot dissolved in $HNO_3$ and HF overnight. The solution was evaporated until dry, and the dissolution process was repeated three times before a final treatment with $HNO_3$. The actinides were then scavenged from the solution using iron hydroxide, separated and purified using ion-exchange chromatography before being electroplated for alpha spectroscopy. Their ratio with our renormalization is 0.013±0.003, and is the lowest of all reported values.

In 2007 Nygren et al.[25] compared gamma ray measurements with inductively coupled plasma sector field mass spectrometry (ICP-SFMS) measurements on chemically-purified samples, obtaining results of 0.0249±0.0047 and 0.0249±0.0001 respectively (the latter is our average for their three samples).

Alpha spectroscopy was used by Belloni et al.[10] in 2011 to determine the plutonium isotopics. No radiochemical separations were done for this measurement, but due to the energy resolution of alpha spectroscopy with conventional detectors, only a [239+240]Pu/[241]Am ratio could be determined.

Fahey et al.[7], Wallace et al.[11] and Bellucci et al.[22] each performed measurements that did not require any chemical pre-processing. Fahey et al. used secondary ion mass spectrometry (SIMS) with a result of 0.0176±0.0006. Wallace et al. used laser ablation inductively coupled plasma mass spectrometry (LA-ICP-MS) at 137 spots on a single sample, obtaining a broad range of results with an average of 0.0208±0.0006. Bellucci et al. analyzed the uranium isotopic composition of trinitite with laser ablation multicollector inductively coupled plasma mass spectrometry (LA-MC-ICP-MS). From the uranium compositions at 12 spots they inferred a [240]Pu/[239]Pu mass ratio of 0.01–0.03. Not only are these techniques independent of chemical concentration, they also are potentially useful to study inhomogeneities within a sample.

A modern analysis by Hanson and Oldham reported in this issue is based on chemical concentration followed by inductively coupled plasma mass spectrometry (ICP-MS).[26] Their ratio, averaged from 14 samples, is 0.0246±0.0003. This may be considered the most reliable measurement of the plutonium isotopic ratio in trinitite to date.

**V. Decay Energy Spectroscopy as a Forensic Tool**

High-resolution calorimetric decay energy spectroscopy (DES) is a recently developed radiometric technique in which a radioactive sample is embedded within the absorber of a low-temperature microcalorimeter.[27,28] For each nuclear decay, the energy of all decay products (alpha particles, gamma-rays, X-rays, electrons, etc.) heats the absorber. Therefore, energy loss in the sample is less of a factor in DES than in alpha spectroscopy, and sample preparation can be simplified. The detector measures 100% of alpha decays, and so the sensitivity of DES is the highest achievable by any radiometric method. Because of the small sample sizes used, separate measurements can be made on sub-samples, a technique which reduces the homogenizing effect of measuring large samples. The temperature change of the absorber is measured by a very sensitive thermometer such as a superconducting transition-edge sensor. For alpha-decaying nuclides ($^{239}$Pu, $^{240}$Pu, $^{241}$Am, etc.), the measured energy corresponds to the unique total nuclear decay energy (Q value). Operation at low temperatures (below 0.1 K) is achieved with commercial cryostat systems and enables ultra-high energy resolution, as good as 1.0 keV FWHM at 5.5 MeV.[28] These characteristics make DES a very sensitive method complementary to traditional forensics techniques for robust, precise isotopic analysis.

DES has been successfully applied to undissolved pure plutonium oxide particles collected from a certified reference material.[28] Trinitite contains only small quantities of plutonium dispersed through a complex matrix and represents a more challenging measurement. To produce small particles for measurement by DES, a 9.16 mg aerodynamic bead (Figure 8A) was ball milled (dry) for 10 minutes. Approximately 50 µg of the resulting powder was embedded in a gold foil by repeatedly folding and pressing the gold foil with pliers to form an absorber for the microcalorimeter detector. A cross section of a representative gold sample with embedded trinitite particles is shown in Figure 8B. The size of trinitite particles varied significantly, but nearly all were below 10 µm in diameter. The assembled microcalorimeter detector is shown in Figure 8C, with the gold absorber attached to the transition-edge sensor chip by an indium bond. The acquired DES spectrum of trinitite, shown in Figure 9, indicates the presence of beta-decaying fission and neutron activation products, along with alpha-decaying actinides. The acquisition time was 19.4 hours, and the total measured alpha activity was 0.04 Bq. A second-order polynomial was used for energy calibration, based on the $^{238}$Pu and $^{239}$Pu peaks. Beta-decaying nuclides cannot be individually quantified, but their aggregate presence is a signature of nuclear fission. The energy resolution is sufficient to identify the presence of $^{239}$Pu, $^{240}$Pu, $^{238}$Pu, and $^{241}$Am, but degraded relative to DES measurements of pure plutonium materials.

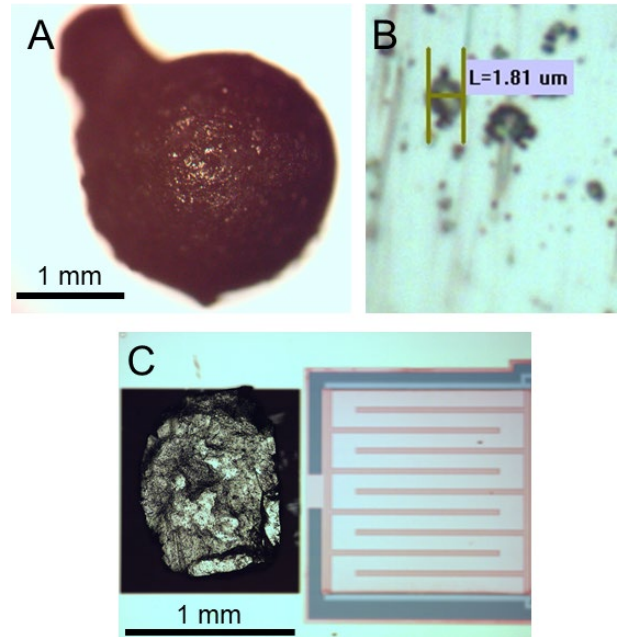

Figure 8. (A) Aerodynamic bead of trinitite used for this measurement. (B) Cross section of representative gold sample shows ball-milled trinitite particles were effectively embedded with a simple folding and pressing process. (C) The gold absorber with embedded particles was attached to the transition-edge sensor with an indium bond. Sample preparation required no dissolution or chemical processing.

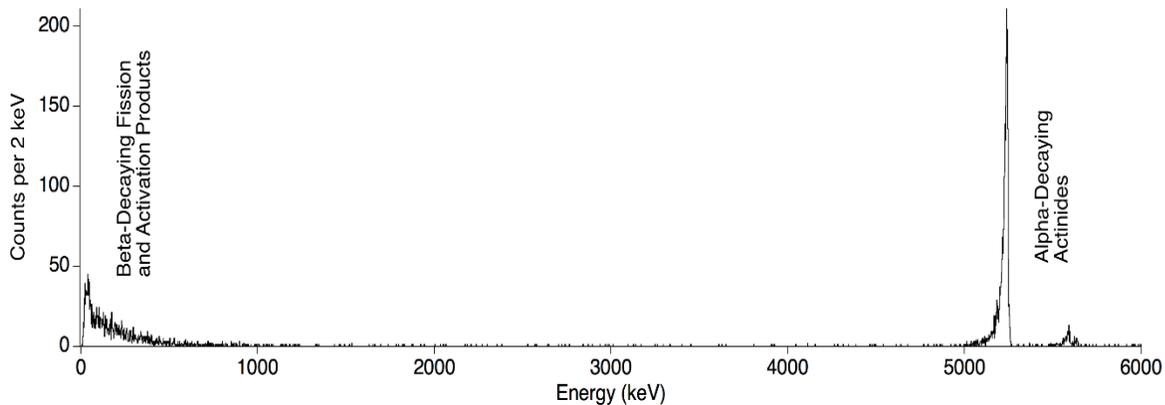

Figure 9. The full DES spectrum of trinitite indicates beta-decaying fission and neutron activation products, primarily below 1 MeV, and alpha-decaying actinides above 5 MeV.

Figure 10 shows the $^{239}$Pu and $^{240}$Pu peaks with a preliminary fit to the data. This region of the spectrum was fit with two Bortels functions (a convolution of a Gaussian and a one-sided exponential tail) constrained to each have the same Gaussian width and tail factor, and maxima constrained to equal to the known difference in Q values between $^{239}$Pu and $^{240}$Pu. The Gaussian component FWHM is 5.1 keV, and the exponential tail factor is 21 keV. Previous DES measurements on pure materials have demonstrated a Gaussian FWHM of 1.0 keV with negligible tailing.[27] We expect that the energy resolution and tailing could be improved by further developing the milling process to produce smaller particles, such that more decay energy would be deposited in the gold absorber matrix rather than the particles of trinitite. Figure 11 shows the region of the spectrum with peaks from $^{238}$Pu and $^{241}$Am. In the measured spectrum, the activity of these two isotopes is too low to merit quantitative analysis, but their presence can be identified.

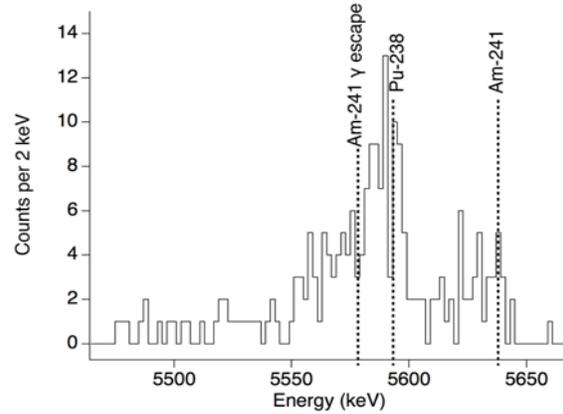

Figure 11. The presence of $^{238}$Pu and $^{241}$Am is apparent in a higher-energy portion of the spectrum. Dashed lines indicate the expected peak locations. The $^{241}$Am 59.5 keV gamma ray escape peak is not seen here but has been observed in other DES measurements.

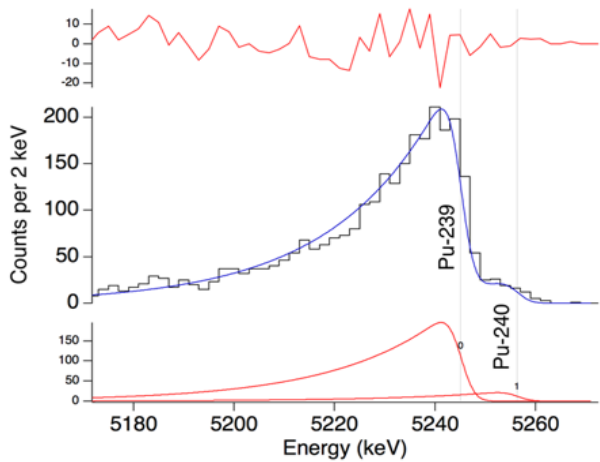

Figure 10. Peaks from $^{239}$Pu and $^{240}$Pu are observed in trinitite. The middle plot shows the fitted spectrum, the top plot shows residuals, and the bottom plot shows the fit components.

Our analysis indicates approximately 16 pg of $^{239}$Pu and approximately 0.5 pg of $^{240}$Pu. Our decay-corrected $^{240}$Pu/$^{239}$Pu mass ratio is 0.030±0.004, with the uncertainty determined mostly by the limited counting statistics of the $^{240}$Pu peak. Our ratio is higher than any of the nine published results shown above, although it is within 1σ of three of them, within 1.5σ for two others, and well within the range of values appearing in Wallace et al..[11] A comparison of all results is given in Figure 12 and Table II. Note that there is no reason to assume all trinitite samples will have identical isotopic ratios; the fireball was not chemically or isotopically homogeneous, so there is no reason to expect the samples will be. The range of results may reflect actual sample variations as well as method-dependent and statistical variations.

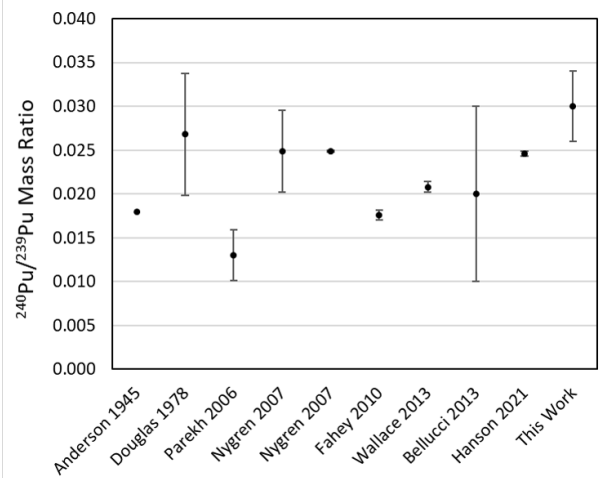

Figure 12. Comparison of reported $^{240}$Pu/$^{239}$Pu mass ratio results.

With an intrinsically low background, DES uncertainty is primarily limited by counting statistics and could be improved with a longer measurement time. It may also be possible to load more trinitite particles into the gold matrix and increase the activity of the sample. It is important to note that there was no cleanup involved in the sample preparation. This is an extremely impure sample, measured with no chemical processing or even dissolution. The plutonium concentration in similar trinitite samples has been measured at less than 1 ppm by mass. This first measurement of its type therefore shows the potential of DES to measure actinide isotopic composition in extremely impure materials with minimal sample preparation.

Table II: $^{240}$Pu/$^{239}$Pu Mass Ratio Observed in Trinitite.

| Reference | Method | Number of Samples (meas.) | Reported Ratio | Minimum | Maximum |
|---|---|---|---|---|---|
| Anderson et al.[23] 1945 | Chem, Fission & Alpha | 1 | 0.018 | | |
| Douglas[24] 1978 | Chem, Alpha & MS | 3 (22) | 0.027±0.007 | 0.0252±0.004 | 0.084±0.003 |
| Parekh et al.[5] 2006 | Chem, Gamma | 1 | 0.0130±0.0029 | | |
| Nygren et al.[25] 2007 | Chem, Gamma | 1 | 0.0249±0.0047 | | |
| Nygren et al.[25] 2007 | Chem, ICP-SFMS | 3 | 0.0249±0.0001 | 0.02486±0.00009 | 0.02493±0.00009 |
| Fahey et al.[7] 2010 | SIMS | 1 | 0.0176±0.0006 | | |
| Wallace et al.[11] 2013 | LA-ICP-MS | 1 (137) | 0.0208±0.0006 | 0.0022±0.0022 | 0.099±0.050 |
| Bellucci et al.[22] 2013 | LA-MC-ICP-MS | 1(12) | 0.02±0.01 | | |
| Hanson et al.[26] 2021 | Chem, ICP-MS | 14 | 0.0246±0.0003 | 0.02425±0.00006 | 0.02508±0.00006 |
| This Work 2021 | DES | 1 | 0.030±0.004 | | |

## VI. Conclusion

The Trinity test site shows clear radiological evidence of a nuclear detonation observable many decades after the event. Radionuclides $^{60}$Co, $^{137}$Cs, $^{152}$Eu, and $^{154}$Eu are plainly visible in the soil, and additionally $^{133}$Ba, $^{239}$Pu, and $^{241}$Am are visible in the mineral trinitite. We see clear differences when comparing trinitite activities to soil activities. Trinitite contains much more $^{241}$Am and has much higher $^{137}$Cs / $^{152}$Eu and $^{60}$Co / $^{152}$Eu ratios than the soil. This may be a result of surface remediation of the site, which removed most of the trinitite but left irradiated subsoil.

We examined atomsite and Kharitonchik samples from the Nevada and Semipalatinsk test sites and found significant differences from Trinitite, suggesting differences in the detonations or processes that created the samples. While the lack of $^{239}$Pu and $^{241}$Am in "Johnnie Boy" atomsite is easily understood, we were surprised by the higher activities in the Kharitonchik sample and have no firm explanation.

Gamma spectroscopy is not a sensitive enough tool to determine Pu isotopics within a small sample of nuclear detonation debris. To accomplish this, we used DES and present the first measurements of nuclear detonation debris with this technique. DES was able to identify $^{238}$Pu, $^{239}$Pu, $^{240}$Pu, and $^{241}$Am and obtain a decay-corrected $^{240}$Pu/$^{239}$Pu mass ratio in a small trinitite bead with a simple, rapid sample preparation process, with no dissolution or chemical processing that could perturb the composition, on a sample three orders of magnitude smaller than that used by Parekh et al..[5] With further development we anticipate that the energy resolution and sensitivity of DES will make it a valuable complementary tool for forensic analysis.

## Acknowledgements

This work was supported in part by the Los Alamos National Laboratory Pathfinder Program. Los Alamos National Laboratory is operated by Triad National Security, LLC, for the National Nuclear Security Administration of the US Department of Energy under Contract No. 89233218CNA000001. Microcalorimeter sensors for decay energy spectroscopy were designed and fabricated by the National Institute of Standards and Technology Quantum Sensors Group. We would like to thank Lisa Blevins of the White Sands Missile Range Public Affairs Office and Omar Juveland, Robert Hermes, David Seagraves, and Byron Ristvet for assistance with samples and measurements.